\documentclass[epj,nopacs]{svjour}
\usepackage{graphics}
\begin{document}
\title{Quarkonia and heavy flavors at the LHC}
\author{Philippe Crochet% etc
\thanks{Philippe.Crochet@clermont.in2p3.fr} %Insert the address here if needed}%
}                     % Do not remove
\institute{Laboratoire de Physique Corpusculaire 
  CNRS/IN2P3
  F-63177 Clermont-Ferrand}
%\email{crochet@clermont.in2p3.fr}
%
\date{Received: date / Revised version: date}
\abstract{Perspectives for measurements of quarkonia and heavy flavors
in heavy ion collisions at the LHC are reviewed.
%
%\PACS{
%      {25.75.-q}{Relativistic heavy-ion collisions}   \and
%      {12.38.Mh}{Quark-gluon plasma} \and
%      {25.75.Nq}{Quark deconfinement, quark-gluon plasma production, and 
%	phase transitions} 
%     } % end of PACS codes
} %end of abstract
\maketitle
\section{Quarkonia and heavy flavors: what is different at the LHC}
\label{widatl}

With a nucleus-nucleus center-of-mass energy nearly 30 times larger
than the one reached at RHIC, the LHC will open a new era for studying
the properties of strongly interacting matter at extreme energy
densities~\cite{Carminati:2004fp}.  One of the most exciting aspects
of this new regime is the abundant production rate of hard probes
which can be used, for the first time, as high statistics probes of
the medium~\cite{Bedjidian:2003gd}.  Futhermore, heavy flavor
measurements at the LHC should provide a comprehensive understanding
of open and hidden heavy flavor production at very low $x$ values,
where strong nuclear gluon shadowing is expected.  The heavy flavor
sector at LHC energies is subject to other significant differences
with respect to SPS and RHIC energies.  First, the large production
rate offers the possibility to use a large variety of observables.
Then, the magnitude of most of the in-medium effects is dramatically
enhanced.  Some of these aspects are discussed hereafter.

\subsection{New observables}

The Table~\ref{qqbar} shows the number of $c\bar{c}$ and $b\bar{b}$
pairs produced in central A-A collisions at SPS, RHIC and LHC.  From
RHIC to LHC, there are 10 times more $c\bar{c}$ pairs and 100 times
more $b\bar{b}$ pairs produced.  Therefore, while at SPS only
charmonium states are experimentally accessible and at RHIC it remains
to be seen how much of the bottom sector can be explored, at the LHC both
charmonia and bottomonia can be used, thus providing powerful probes
for Quark Gluon Plasma (QGP) studies.  In fact, since the
$\Upsilon(1S)$ state only dissolves significantly above the critical
temperature~\cite{Digal:2001ue}, at a value which might only be
reachable above that of RHIC, the spectroscopy of the $\Upsilon$
family at the LHC should reveal unique characteristics of the
QGP~\cite{Gunion:1996qc}.  In addition to the centrality dependence of
the $\Upsilon$ yield, the study of the $\Upsilon^\prime/\Upsilon$
ratio versus transverse momentum ($p_{\rm T}$) is believed to be of
crucial interest~\cite{Gunion:1996qc} (see below).

\begin{table}[ht]
\centering
\caption{Number of $c\bar{c}$ and $b\bar{b}$ pairs produced in central
heavy-ion collisions ($b=0$) at SPS (Pb-Pb), RHIC (Au-Au), and LHC
(Pb-Pb) energies. $b\bar{b}$ production is negligible at the SPS.}
\label{qqbar}       % Give a unique label
\begin{tabular}{lccc}
\hline\noalign{\smallskip}
& SPS & RHIC & LHC  \\
\noalign{\smallskip}\hline\noalign{\smallskip}
N($c\bar{c}$) & 0.2 & 10 & 130 \\
N($b\bar{b}$) & -- & 0.05 & 5 \\
\noalign{\smallskip}\hline
\end{tabular}
\end{table}
On the other hand, studies with open heavy flavors also benefit from
high statistics measurements.  In particular, as shown in the
following, the reconstruction of the $p_{\rm T}$ distribution of $D^0$
mesons in the hadronic channel should provide valuable information on
in-medium induced $c$ quark energy loss.

\subsection{Large quarkonium nuclear absorption}

Charmonium measurements at the SPS have shown that a detailed
understanding of the normal nuclear absorption is mandatory in order
to reveal any anomalous suppression behavior~\cite{louis}.  According
to Ref.~\cite{Bedjidian:2003gd}, the following observations can be
made:
\begin{itemize}
\item the J/$\psi$ nuclear absorption in central Pb-Pb collisions is
  two times larger at the LHC than at the SPS;
\item the J/$\psi$ nuclear absorption in central Ar-Ar collisions at
the LHC is similar to the one in central Pb-Pb collisions at the SPS;
\item the $\Upsilon$ nuclear absorption in central Pb-Pb collisions at
the LHC is similar to the J/$\psi$ nuclear absorption in central Pb-Pb
collisions at the SPS.
\end{itemize}

\subsection{Large resonance dissociation rate}

It has been realized that, in addition to the normal nuclear
absorption, the interactions with comoving hadrons and the melting by
color screening, quarkonia can also be significantly destroyed by
gluon ionization~\cite{Xu:1995eb}.  Since this mechanism results from
the presence of quasi-free gluons, it starts being effective for
temperatures above the critical temperature but not necessarily above
the resonance dissociation temperature by color screening.  Recent
estimates~\cite{Bedjidian:2003gd} (see Ref.~\cite{Blaschke:2004dv} for
an update) of the quarkonium dissociation cross-sections show that
none of the J/$\psi$ mesons survives the deconfined phase at the LHC
and that about 80\,\% of the $\Upsilon$ are destroyed.  Significant
information about the initial temperature and lifetime of the QGP
should be extracted from the $\Upsilon$ suppression pattern.

\subsection{Large charmonium secondary production}

An important yield of secondary charmonia is expected from $B$ meson
decays~\cite{Eidelman:2004wy}, $D\overline{D}$
annihilation~\cite{Ko:1998fs}, statistical
hadronization~\cite{Braun-Munzinger:2000px} and kinetic
recombination~\cite{Thews:2000rj}.  Contrary to the two first
processes, the two last ones explicitly assume the formation of a
deconfined medium.  The underlying picture is that charmonium
resonances form by coalescence of free $c$ and $\bar{c}$ quarks in the
QGP~\cite{Thews:2000rj} or at the hadronization
stage~\cite{Braun-Munzinger:2000px}.  According to these models, the
QGP should lead to an increase of the J/$\psi$ yield versus
centrality, roughly proportional to ${\rm N}^2(c\bar{c})$, instead of
a suppression.  Due to the large number of $c\bar{c}$ pairs produced
in central heavy ion collisions at the LHC, these models predict a
spectacular enhancement of the J/$\psi$ yield; up to a factor 100
relative to the primary production
yield~\cite{Bedjidian:2003gd,Andronic:2003zv}.  Although the
statistical accuracy of the present RHIC data cannot confirm or rule
out such mechanisms~\cite{robert}, it is interesting to extrapolate
from secondary charmonium production at RHIC to secondary bottomonium
production at the LHC.  Indeed, the expected multiplicity of
$b\bar{b}$ pairs at the LHC is roughly equal to the expected
multiplicity of $c\bar{c}$ pairs at RHIC (Table~\ref{qqbar}).
Therefore, if secondary production of charmonia is observed at RHIC,
it is conceivable to expect the same formation mechanism for
bottomonium states at the LHC.

\subsection{Complex structure of dilepton yield}

\begin{figure}
\begin{center}
  \resizebox{0.35\textwidth}{!}{\includegraphics{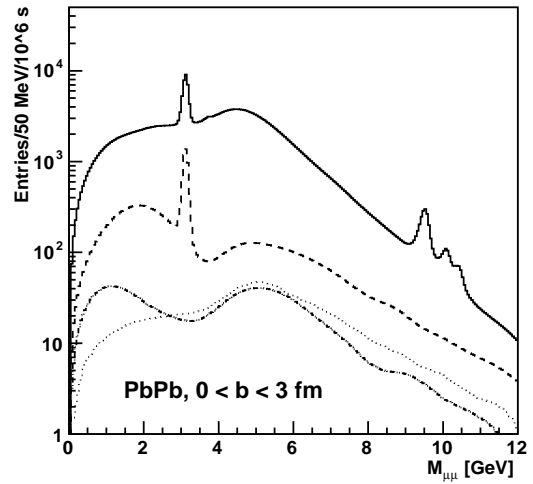}}
\end{center}
\caption{Invariant mass spectra of dimuons produced in central
    ($b<3$~fm) Pb-Pb collisions in the ALICE forward muon
    spectrometer~\cite{smbat}, with a $p_{\rm T}$ cut of 2~GeV/$c$
    applied to each single muon. The lines correspond to: like-sign
    correlated dimuons from bottom (dotted); unlike-sign correlated
    dimuons from charm (dash-dotted) and from bottom (dashed); 
    unlike-sign correlated and unlike-sign non-correlated pairs (solid).}
\label{smbat2}       
\end{figure}

The dilepton mass spectrum at the LHC exhibits new features,
illustrated in Fig.~\ref{smbat2}.  It can be seen that, with a low
$p_{\rm T}$ threshold of around 2~GeV/$c$ on the decay leptons,
unlike-sign dileptons from bottom decay dominate the dilepton
correlated component over all the mass range.  These dileptons have
two different origins.  In the high invariant mass region, each lepton
comes from the direct decay of a $B$ meson (the so-called $BB$-diff
channel).  In the low invariant mass region, both leptons come from
the decay of a single $B$ meson via a $D$ meson (the so-called
$B$-chain channel).  Next to leading order processes, such as gluon
splitting, also populate significantly the low mass dilepton spectrum
due to their particular kinematics.  Then, as discussed in more detail
below, a substantial fraction of the J/$\psi$ yield arises from bottom
decays.  Finally, a sizeable yield of like-sign correlated dileptons
from bottom decays is present.  This contribution arises from the
peculiar decay chain of $B$ mesons and from $B$ meson oscillations
(see below).  Its yield could be even larger than the yield of
unlike-sign correlated dileptons from charm.

\section{The LHC heavy ion program}

The LHC will be operated several months per year in pp mode and
several weeks in heavy-ion mode.  The corresponding effective time for
rate estimates is $10^7$~s for pp and $10^6$~s for heavy-ion
operation.  As described in Ref.~\cite{Carminati:2004fp}, the
``heavy-ion runs'' include, during the first five years of operation,
one Pb-Pb run at low luminosity, two Pb-Pb runs at high luminosity,
one p-A run and one light-ion run.  In the following years different
options will be considered, depending on the first results.  Three of
the four LHC experiments are expected to take heavy-ion data.

\subsection{ALICE}

ALICE (A Large Ion Collider Experiment) is the only LHC experiment
dedicated to the study of nucleus-nucleus collisions~\cite{ALICEWEB}.
The detector is designed to cope with large charged particle
multiplicities which, in central Pb-Pb collisions, are expected to be
between 2000 and 8000 per unit rapidity at mid rapidity.  The detector
consists of a central barrel ($|\eta|<0.9$), a forward muon
spectrometer $(2.5<\eta<4$) and several forward/backward and central
small acceptance detectors.  Heavy flavors will be measured in ALICE
through the electron channel and the hadron channel in the central
barrel as well as through the muon channel in the forward region.
Note that, contrary to the other LHC experiments, ALICE will be able
to access most of the signals down to very low $p_{\rm T}$.

\subsection{CMS}

CMS (Compact Muon Solenoid)~\cite{CMSWEB} is designed for high $p_{\rm
T}$ physics in pp collisions but has a strong heavy ion
program~\cite{cms}.  This program includes jet reconstruction,
quarkonia measurements (in the dimuon channel) and high mass dimuon
measurements.  The detector acceptance, for quarkonia measurements,
ranges from $-2.5$ to 2.5 in $\eta$, with a $p_{\rm T}$ threshold of
3~GeV/$c$ on single muons.  Such a $p_{\rm T}$ cut still allows the
reconstruction of $\Upsilon$ states down to $p_{\rm T} = 0$ but limits
J/$\psi$ measurements to high $p_{\rm T}$.

\subsection{ATLAS}

Like CMS, ATLAS (A Toroidal LHC ApparatuS)~\cite{ATLASWEB} is designed
for pp physics.  The detector capabilities for heavy ion physics have
been investigated recently~\cite{atlas}.  As far as heavy flavors are
concerned, the physics program will focus on measurements of $b$-jets
and $\Upsilon$.  The detector acceptance for muon measurements is
large in $\eta$ ($|\eta|<2.4$) but, like CMS, is limited to high
$p_{\rm T}$.

\section{Selected physics channels}
\subsection{Quarkonia}
\subsubsection{Centrality dependence of resonance yields}

The centrality dependence of the quarkonium yield, in the $\mu\mu$
channel, has been simulated in the ALICE detector.  From the results,
displayed in Table~\ref{smbatTable}, the following comments can be
made.  The statistics of J/$\psi$ events is large and should allow for
narrower centrality bins.  The $\psi^\prime$ measurement is rather
uncertain, because of the small signal to background ratio (S/B).  The
$\Upsilon$ and $\Upsilon^\prime$ statistics and significance are quite
good and the corresponding S/B ratios are almost always greater
than~1.  On the other hand, the $\Upsilon^{\prime\prime}$ suffers from
limited statistics.  The resonances will also be measured in the
dielectron channel in ALICE~\cite{TRDTP}, and in the dimuon channel in
CMS~\cite{cms} and ATLAS~\cite{atlas}, providing consistency
cross-checks and a nice complementarity in acceptance.  A recent
study~\cite{sudhir} demonstrated the capabilities of ALICE to
measure the resonance azimuthal emission angle with respect to the
reaction plane.  Such measurements are of particular importance given
the latest RHIC results on open charm elliptic
flow~\cite{Kelly:2004qw}.

\begin{table}[ht]
\caption{Preliminary yield (S), signal over background (S/B) and
significance (${\rm S}/\sqrt{\rm S+B}$) for quarkonium resonances
measured versus centrality in the ALICE forward muon
spectrometer~\cite{smbat}.  The input cross-sections are taken from
Ref.~\cite{Bedjidian:2003gd}.  Shadowing is taken into account.  Any
other suppression or enhancement effects are not included.  The
numbers correspond to one month of Pb-Pb data taking and are extracted
with a 2$\sigma$ mass cut.}
\label{smbatTable}       % Give a unique label
\begin{tabular}{lllllll}
\hline\noalign{\smallskip}
& $b$ (fm) & 0-3 & 3-6 & 6-9 & 9-12 & 12-16  \\
%& $\epsilon$ $({\rm GeV/fm}^3)$ & 32 & 30 & 28 & 16 & 5 \\
\noalign{\smallskip}\hline\noalign{\smallskip}
         & S $(\times 10^3)$        & 86.48 & 184.6 & 153.3 & 67.68 & 10.46 \\
J/$\psi$ & S/B                      & 0.167 & 0.214 & 0.425 & 1.237 & 6.243 \\
         & ${\rm S}/\sqrt{\rm S+B}$ & 111.3 & 180.4 & 213.8 & 193.4 & 94.95 \\
\noalign{\smallskip}\hline\noalign{\smallskip}
         & S $(\times 10^3)$        & 1.989 & 4.229 & 3.547 & 1.565 & 0.24 \\
$\psi^\prime$ & S/B                 & 0.009 & 0.011 & 0.021 & 0.063 & 0.273 \\
         & ${\rm S}/\sqrt{\rm S+B}$ & 4.185 & 6.902 & 8.604 & 9.641 & 7.171 \\ 
\noalign{\smallskip}\hline\noalign{\smallskip}
           & S $(\times 10^3)$      & 1.11  & 2.376 & 1.974 & 0.83  & 0.118 \\
$\Upsilon$ & S/B                    & 2.084 & 2.732 & 4.31  & 7.977 & 12.01 \\
           & ${\rm S}/\sqrt{\rm S+B}$ & 27.39 & 41.71 & 40.03 & 27.16 & 10.42\\
\noalign{\smallskip}\hline\noalign{\smallskip}
       & S $(\times 10^3)$        & 0.305 & 0.653 & 0.547 & 0.229 & 0.032 \\
$\Upsilon^\prime$ & S/B           & 0.807 & 1.043 & 1.661 & 2.871 & 4.319 \\
       & ${\rm S}/\sqrt{\rm S+B}$ & 11.68 & 18.26 & 18.48 & 13.02 & 5.077 \\ 
\noalign{\smallskip}\hline\noalign{\smallskip}
& S $(\times 10^3)$        & 0.175 & 0.376 & 0.312 & 0.13  & 0.019 \\
$\Upsilon^{\prime\prime}$ & S/B & 0.566 & 0.722 & 1.18  & 1.936 & 3.024 \\
& ${\rm S}/\sqrt{\rm S+B}$ & 7.951 & 12.55 & 13    & 9.274 & 3.73 \\       
\noalign{\smallskip}\hline
\end{tabular}
% Or use
%\vspace*{5cm}  % with the correct table height
\end{table}

\subsubsection{$\Upsilon^\prime/\Upsilon$ ratio versus $p_{\rm T}$}

\begin{figure*}
\begin{center}
  \resizebox{0.74\textwidth}{!}{\includegraphics{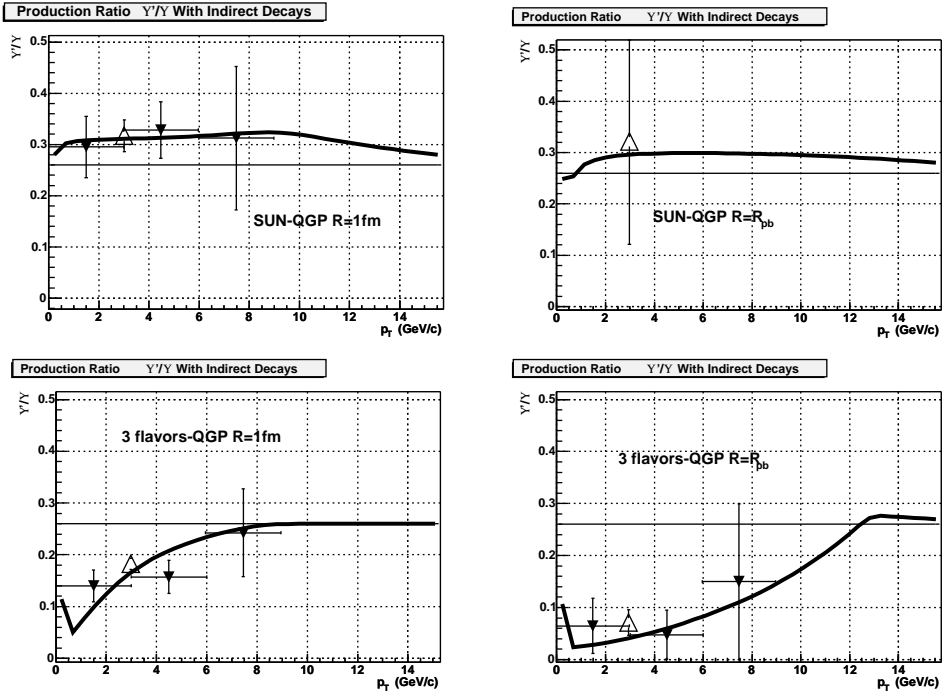}}
\end{center}
  \caption{$\Upsilon^\prime/\Upsilon$ ratio versus $p_{\rm T}$ for two
    different QGP models with different system
    sizes~\cite{ericTHESIS}.  The solid curves correspond to the
    ``theoretical ratios''.  The triangles show the expected
    measurements with the ALICE forward muon spectrometer in one month
    of central (10\,\%) Pb-Pb data taking (the open triangles
    correspond to the $p_{\rm T}$ integrated ratios).  Error bars are
    of statistical origin only.  The horizontal solid lines show the
    expected value of the ratio in pp collisions.  More details on the
    ingredients used in the different scenarios can be found in
    Ref~\cite{Gunion:1996qc}.}
  \label{ericFIG}       
\end{figure*}

The $p_{\rm T}$ dependence of resonance suppression was recognized
very early as a relevant observable to probe the characteristics of
the deconfined medium~\cite{Blaizot:1987ha}.  Indeed, the $p_{\rm T}$
suppression pattern of a resonance is a consequence of the competition
between the resonance formation time and the QGP temperature, lifetime
and spatial extent.  However, quarkonium suppression is known to
result not only from deconfinement but also from nuclear effects like
shadowing and absorption.  In order to isolate pure QGP effects, it
has been proposed to study the $p_{\rm T}$ dependence of quarkonium
ratios instead of single quarkonium $p_{\rm T}$ distributions.  By
doing so, nuclear effects are washed out, at least in the $p_{\rm T}$
variation of the ratio\footnote{Using ratios has the additional
advantage that systematical detection inefficiencies cancel out to
some extent.}.  Following the arguments of Ref.~\cite{Gunion:1996qc},
the capabilities of the ALICE muon spectrometer to measure the $p_{\rm
T}$ dependence of the $\Upsilon^\prime/\Upsilon$ ratio in central
(10\,\%) Pb-Pb collisions have been recently
investigated~\cite{ericTHESIS}.  Two different QGP models with
different system sizes were considered.  The results of the
simulations (Fig.~\ref{ericFIG}) show that, with the statistics
collected in one month of data taking, the measured
$\Upsilon^\prime/\Upsilon$ ratio exhibit a strong sensitivity to the
characteristics of the QGP.  Note that in the scenario of the upper
right panel of Fig.~\ref{ericFIG} the expected suppression is too
large for any measurement beyond the $p_{\rm T}$ integrated one.

\subsubsection{Secondary J/$\psi$ from bottom decay}

A large fraction of the J/$\psi$ yield arises from the decay of $B$
mesons.  The ratio ${\rm N}(b\bar{b}\rightarrow J/\psi)/{\rm N}({\rm
direct}~J/\psi)$ can be determined as follows.  The number of directly
produced J/$\psi$ in central (5\,\%) Pb-Pb collisions is
0.31~\cite{Bedjidian:2003gd}\footnote{Including shadowing and no
feed-down from higher states.}.  The corresponding number of
$b\bar{b}$ pairs (with shadowing) amounts to
4.56~\cite{Bedjidian:2003gd}.  The $b\rightarrow {\rm J}/\psi$
branching ratio is $1.16\pm0.10\,\%$~\cite{Eidelman:2004wy}.  Therefore
${\rm N}(b\bar{b}\rightarrow {\rm J}/\psi)/{\rm N}({\rm direct}~{\rm
J}/\psi) = 34\,\%$ in $4\pi$.  These secondary J/$\psi$ mesons from
$b$ decays, which are not QGP suppressed, must be subtracted from the
measured J/$\psi$ yield prior to J/$\psi$ suppression
studies\footnote{In addition, 1.5\,\% of $B$ mesons decay into
$\chi_{c1}(1P)$ which subsequently decay into $\gamma$J$/\psi$ with a
31\,\% branching ratio~\cite{Eidelman:2004wy}.}.  They can further be
used in order to measure the $b$ cross-section in pp
collisions~\cite{Acosta:2004yw}, to estimate shadowing in p-A
collisions and to probe the medium induced $b$ quark energy loss in
A-A collisions.  Indeed, it has been shown~\cite{Lokhtin:2001nh} that
the $p_{\rm T}$ and $\eta$ distributions of those J/$\psi$ exhibit
pronounced sensitivity to $b$ quark energy loss.  In addition, a
comparison between high mass dileptons and secondary J/$\psi$
distributions could clarify the nature of the energy
loss~\cite{Lokhtin:2001nh}.

Due to the large life-time of $B$ mesons, J/$\psi$ from bottom decay
is the only source of J/$\psi$ not coming from the primary
vertex\footnote{J/$\psi$ from statistical hadronization, kinetic
recombination and $D\overline{D}$ annihilation are usually quoted as
secondary J/$\psi$ but they originate from the primary vertex.}.  The
best way to identify them is, therefore, to reconstruct the invariant
mass of dileptons with displaced vertices i.e.\ with impact parameter,
$d0$, above some threshold.  Simulations have shown that such
measurements can successfully be performed with dielectrons measured
in the central part of ALICE using the ITS, the TPC and the
TRD~\cite{TRDTP} and with dimuons in CMS~\cite{Lokhtin:2001nh}, thanks
to the excellent spatial resolution of the inner tracking devices of
these experiments.  It should also be possible to disentangle the two
sources of J/$\psi$ from the slopes of the overall measured J/$\psi$
$p_{\rm T}$ distributions since primary J/$\psi$ have a harder
spectrum~\cite{TRDTP}.  Finally, a recent study~\cite{andreas} has
demonstrated the possibility to isolate J/$\psi$ from bottom decay in
pp collisions, without secondary vertex reconstruction, by triggering
on three muon events in the ALICE forward muon spectrometer.  Indeed,
in standard (dimuon) pp events, the J/$\psi$ peak contains 85\,\% of
primary J/$\psi$ and 15\,\% of J/$\psi$ from $B$ meson decays.  The
situation is totally inverted in tri-muon events because a
$B\overline{B}$ pair can easily produce many decay leptons.  In such
events the J/$\psi$ peak contains 85\,\% of secondary J/$\psi$ from
bottom decay and 15\,\% of direct J/$\psi$~\cite{andreas}.  It is
obvious that this analysis technique becomes less and less efficient
as the track multiplicity increases.  Nevertheless, it could still be
performed for light-ion systems.

\subsection{Open heavy flavors}

\subsubsection{Open bottom from single leptons with displaced vertices}

As mentionned above, the $d0$ distributions of leptons from heavy
meson decays exhibit a significantly large tail because heavy mesons
have a larger life-time than other particles decaying into leptons.
Therefore, inclusive measurements of open heavy flavors can be
achieved from the identification of the semi-leptonic decay of heavy
mesons~\cite{TRDTP}.  Recent simulation studies~\cite{rosario}
performed with the ALICE central detectors show that with
$d0>180~\mu{\rm m}$ and $p_{\rm T}>2$~GeV/$c$, the monthly expected
statistics of electrons from $B$ decays in central Pb-Pb collisions is
$5\cdot 10^4$ with a contamination of only 10\,\%, mainly coming from
charm decays.  The deconvolution of $d0$ distributions by imposing
different $p_{\rm T}$ cuts should allow charm measurements as well.
Furthermore, such analyses should give access to the $p_{\rm T}$
distribution of $D$ and $B$ mesons by exploiting the correlation
between the $p_{\rm T}$ of the decay lepton and that of its
parent~\cite{TRDTP}.

\subsubsection{Open bottom from single muons and unlike-sign dimuons}

\begin{figure}[ht]
\begin{center}
  \resizebox{0.40\textwidth}{!}{\includegraphics{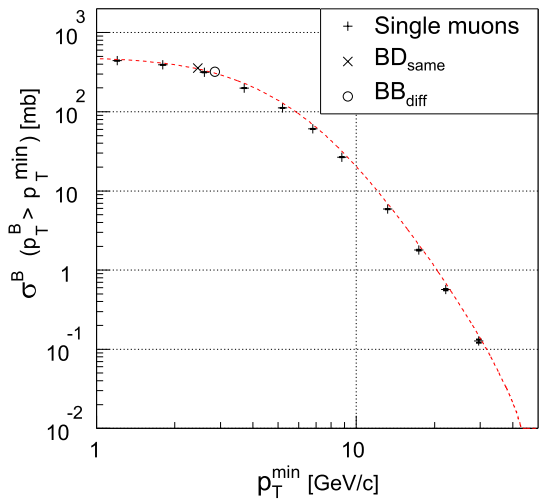}}
\end{center}
 \caption{Differential $B$ hadron inclusive production cross-section
   in the most central (5\,\%) Pb-Pb collisions~\cite{rachid}.
   Measurements from unlike-sign dimuons at low and high mass and from
   single muons (symbols) are compared to the input distribution
   (curve).  Statistical errors (not shown) are negligible.}
  \label{rachidFIG}       
\end{figure}

The possibility to measure the differential $B$ hadron inclusive
production cross-section in central Pb-Pb collisions at the LHC has
recently been investigated by means of analyses similar to the ones
performed in p$\bar{\rm p}$ collisions at the Tevatron.  This study is
based on unlike-sign dimuon mass and single muon $p_{\rm T}$
distributions measured with the ALICE forward muon
spectrometer~\cite{rachid}.  The principle is first to apply a low
$p_{\rm T}$ threshold on single muons in order to reject background
muons (mainly coming from charm decays) and therefore to maximize the
$b$ signal significance.  Then, fits are performed to the total
(di)muon yield with fixed shapes for the different contributing
sources and the bottom amplitude as the only free parameter.  The $B$
hadron production cross-section is then obtained after corrections for
decay kinematics and branching ratios and muon detection acceptance
and efficiencies.  This allows to extract the signal over a broad
range in $p_{\rm T}$ (Fig.~\ref{rachidFIG}).  A large statistics is
expected~\cite{rachid} thus allowing detailed investigations on $b$
quark production mechanisms and in-medium energy loss.  On the other
hand, such a measurement, which can be performed for different
centrality classes, provides the most natural normalization for
$\Upsilon$ suppression studies.

\subsubsection{Open bottom from like-sign dileptons}

As shown in Fig.~\ref{smbat2}, a sizable fraction of like-sign
correlated dileptons arise from the decay of $B$ mesons.  These
dileptons have two different origins:
\begin{itemize}
\item{The first decay generation of $B$ mesons contains $\sim 10\,\%$ of
primary leptons and a large fraction of $D$ mesons which decay
semi-leptonically with a branching ratio of $\sim 12\,\%$.  Therefore a
$B\overline{B}$ pair is a source of like-sign correlated dileptons via
channels like:\\
\hspace*{0.5cm}$B^+$ $\rightarrow$ $\overline{D}^0$ $e^+$ $\nu_e$,
$\overline{D}^0$ $\rightarrow$ $e^-$ anything\\
\hspace*{0.5cm}$B^-$ $\rightarrow$ $D^0$ $\pi^-$, $D^0$ $\rightarrow$
$e^+$ anything\\ where the $B^+B^-$ pair produces a correlated
$e^+e^+$ pair in addition to the two correlated $e^+e^-$ pairs;}
\item{The two neutral $B^0\overline{B}^0$ meson systems
$B^0_d\overline{B}^0_d$ and $B^0_s\overline{B}^0_s$ undergo the
phenomenon of particle-antiparticle mixing (or oscillation).  The
mixing parameters\footnote{Time-integrated probability that a produced
$B^0_d$ ($B^0_s$) decays as a $\overline{B}^0_d$ ($\overline{B}^0_s$)
and vice versa.}  are $\chi_d~= 0.17$ and $\chi_s\ge
0.49$~\cite{Eidelman:2004wy}.  Therefore, a $B^0_{d}\overline{B}^0_d$
($B^0_s\overline{B}^0_s$) pair produces, in the first generation of
decay leptons, $70\,\%$ ($50\,\%$) of unlike-sign correlated lepton pairs
and $30\,\%$ ($50\,\%$) of like-sign correlated lepton pairs.}
\end{itemize}
This component is accessible experimentally from the subtraction of
so-called event-mixing spectrum from the like-sign
spectrum~\cite{Crochet:2001qd}.  The corresponding signal is a
reliable measurement of the bottom cross-section since i) $D$ mesons
do not oscillate~\cite{Eidelman:2004wy} and ii) most (if not all)
leptons from the second generation of $D$ meson decay can be removed
by a low $p_{\rm T}$ threshold of about 2~GeV/$c$.

\subsubsection{Hadronic charm}

In the central part of ALICE, heavy mesons can be fully reconstructed
from their charged particle decay products in the ITS, TPC and
TOF~\cite{Dainese:2003zu}.  Not only the integrated yield, but also
the $p_{\rm T}$ distribution can be measured.  The most promising
decay channel for open charm detection is the $D^0 \rightarrow
K^-\pi^+$ decay (and its charge conjugate) which has a branching ratio
of 3.8\,\% and $c\tau=124~\mu{\rm m}$.  The expected rates (per unit
of rapidity at mid rapidity) for $D^0$ (and $\overline{D}^0$) mesons,
decaying in a $K^\mp\pi^\pm$ pair, in central (5\,\%) Pb-Pb at
$\sqrt{s}=5.5~{\rm TeV}$ and in pp collisions at $\sqrt{s}=14~{\rm
TeV}$, are $5.3\cdot 10^{-1}$ and $7.5\cdot 10^{-4}$ per event,
respectively.  The selection of this decay channel allows the direct
identification of the $D^0$ particles by computing the invariant mass
of fully-reconstructed topologies originating from displaced secondary
vertices.  The expected statistics are $\sim 13\,000$ reconstructed
$D^0$ in $10^7$ central Pb-Pb collisions and $\sim 20\,000$ in $10^9$
pp collisions.  The significance is larger than 10 for up to about
$p_{\rm T}=10$~GeV/$c$ both in Pb-Pb and in pp collisions.  The
cross section can be measured down to $p_{\rm T} = 1$~GeV/$c$ in Pb-Pb
collisions and down to almost $p_{\rm T} = 0$ in pp collisions.

\begin{figure}[ht]
\begin{center}
  \resizebox{0.46\textwidth}{!}{\includegraphics{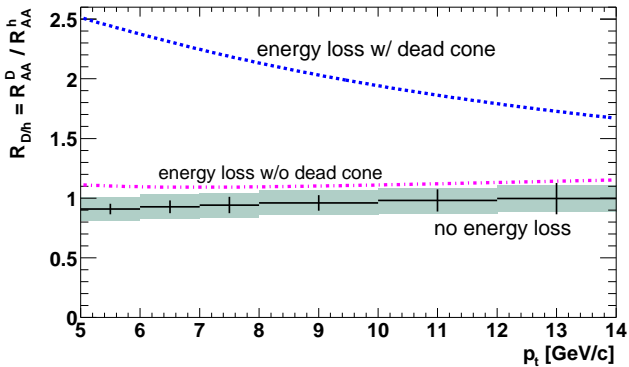}}
\end{center}
  \caption{Ratio of the nuclear modification factors for $D^0$ mesons
    and for charged (non-charm) hadrons with and without energy loss
    and dead cone effect~\cite{Dainese:2003wq}.  Errors corresponding
    to the case ``no energy loss'' are reported.  Vertical bars and
    shaded areas correspond to statistical and systematic errors,
    respectively.}
  \label{andrea2}       
\end{figure}

The reconstructed $D^0$ $p_{\rm T}$ distributions can be used to
investigate the energy loss of $c$ quarks by means of the nuclear
modification factor $R_{\rm
AA}^{D^0}$~\cite{Dainese:2003zu,Dainese:2003wq}.  Even more
interesting is the ratio of the nuclear modification factors of $D^0$
mesons and of charged (non-charm) hadrons ($R_{D/h}$) as a function of
$p_{\rm T}$.  Apart from the fact that many systematic uncertainties
on $R_{\rm AA}^{D^0}$ cancel out with the double ratio, $R_{D/h}$
offers a powerful tool to investigate and quantify the so-called dead
cone effect (Fig.~\ref{andrea2}).

\subsubsection{Electron-muon coincidences}

The semi-leptonic decay of heavy mesons involves either a muon or an
electron.  Therefore, the correlated $c\bar{c}$ and $b\bar{b}$
cross-sections can be measured in ALICE from unlike-sign electron-muon
pairs where the electron is identified in the central part and the
muon is detected in the forward muon spectrometer.  The $e\mu$ channel
is the only leptonic channel which gives a direct access to the
correlated component of the $c\bar{c}$ and $b\bar{b}$ pairs.  Indeed,
in contrast to $e^+e^-$ and $\mu^+\mu^-$ channels, neither a
resonance, nor direct dilepton production, nor thermal production can
produce correlated $e\mu$ pairs.  Within ALICE, the $e\mu$ channel has
the additional advantage that the rapidity distribution of the
corresponding signal extends from $\sim 1$ to $\sim 3$, therefore
bridging the acceptances of the central and the forward parts of the
detector~\cite{Lin:1998bd}.  Electron-muon coincidences have already
been successfully measured in pp collisions at $\sqrt{s}=60~{\rm
GeV}$~\cite{Chilingarov:1979ur} and in p-nucleus collisions at
$\sqrt{s}=29~{\rm GeV}$~\cite{Akesson:1996wf}.  Preliminary
simulations have shown the possibility, with ALICE, to measure the
correlated $e\mu$ signal after appropriate background
subtraction~\cite{MUONTDR}.

\section{Summary}

The heavy flavor sector will bring fantastic opportunities for
systematic explorations of the dense partonic system formed in heavy
ion collisions at the LHC through a wide variety of physics channels.
In addition to the channels discussed here, further exciting
possibilities should be opened with, for example, charmed baryons,
high mass dileptons, quarkonia polarization and dilepton correlations.

\section*{Acknowledgments}
I am grateful to A.~Andronic, M.~Bedjidian, A.~Dainese, S.~Grigoryan,
R.~Guernane, G.~Martinez and A.~Morsch for their help in preparing
this paper.

%%\cite{Dainese:2004rh}
%\bibitem{Dainese:2004rh}
%A.~Dainese  [ALICE Collaboration],
%%``Heavy flavours in heavy-ion collisions at the LHC: ALICE performance,''
%arXiv:nucl-ex/0405008.
%%%CITATION = NUCL-EX 0405008;%%
%----------------------------------------------

\end{document}